\documentstyle[amssymb,aps,preprint,epsf,psfig]{revtex}

\begin{document}
\title{Nonequilibrium magnetotransport through a quantum dot: 
An interpolative perturbative approach}
\author{A. A. Aligia}
\address{Comisi\'{o}n Nacional de Energ\'{\i }a At\'{o}mica,\\
Centro At\'{o}mico Bariloche and Instituto Balseiro, 8400 Bariloche,\\
Argentina }
\maketitle

\begin{abstract}
We study the differential conductance, spectral density and
magnetization, for a quantum dot coupled to two conducting leads as a
function of bias voltage $V_{ds}$, magnetic field $B$ and temperature $T$.
The system is modeled with the Anderson model solved using a spin dependent
interpolative perturbative approximation in the Coulomb repulsion $U$ that
conserves the current. For large enough magnetic field, the differential
conductance as a function of bias voltage shows split peaks. This splitting
is larger than the corresponding splitting in the spectral density of states 
$\rho _{d}(\omega )$, in agreement with experiment.
\end{abstract}





\section{Introduction}

Electron transport through nanoscale quantum dots (QD's) has been a subject
of great interest in the last years. A QD consists of a confined droplet of
electrons and it has been predicted that the system behaves as a
single-electron transistor described by the Anderson model.\cite{glaz,ng}
Experiments with one QD in the linear response regime \cite
{gold1,cro,gold2,wiel} displayed clearly the Kondo physics in this almost
ideal ``one-impurity'' system and confirmed the predictions based on the
Anderson model.\cite{glaz,ng,chz} The unitary limit of ideal transmittance
has been reached.\cite{wiel} Calculations using the accurate Wilson's
numerical renormalization group (NRG) have shown a good agreement with these
experiments.\cite{izum} The effects of a magnetic field and the possibility
to use transport through a QD\ or a site coupled to a QD as a spin filter
have also been addressed theoretically,\cite{rech,costi,torio,lady}
including NRG results.\cite{costi}

In contrast to the linear response regime, the experimental situation in
which a finite bias voltage between drain and source $V_{ds}$ is applied,%
\cite{def,kogan,zum,ama,nyg,letu} i.e. the nonequilibrium case, is much
harder to treat theoretically. The most accurate techniques used to study
the impurity Anderson and Kondo models at equilibrium,\cite{hews} cannot be
easily extended to the non-equilibrium situation. The NRG has not been
implemented out of equilibrium. A formalism to generalize exact Bethe ansatz
results for the nonequilibrium case has been proposed recently,\cite{mehta}
but its application to the Kondo model requires further elaboration. Some
results based on integrabilty but not exact have been presented.\cite{konik}
Also a Kondo model has been solved exactly for particular parameters.\cite
{schi} The non-crossing approximation (NCA), which has been successfully
used to calculate thermodynamic properties of generalized Anderson models
above or near the Kondo temperature $T_{K}$,\cite{bick,monnier,nelson} has
also been used for nonequilibrium situations.\cite{meirn,wing,ros1}
Unfortunately, at low temperatures this method fails to fulfill Fermi liquid
relations in the equilibrium case,\cite{mull} and the spectral density
presents a spurious peak at the Fermi level in the presence of a magnetic
field $B$.\cite{wing} A likely cause of this peak is given by Moore and Wen,
who study the splitting of the Kondo resonance with $B$ from the
Bethe-ansatz solution at equilibrium.\cite{moore} Slave-boson mean-field
approximations have been used successfully in the equilibrium case,\cite
{lady,kang,hu,ihm,lin,revi} but they rely on the minimization of the
free energy, which is not at a minimum in the nonequilibrium case. 
Nevertheless, these approximations have been used for problems out of 
equilibrium.\cite{agua,dong}
A relatively simple but accurate approach based on perturbation theory is the
local moment approach,\cite{logan,logan2,vid} but its extension to the
nonequilibrium case is not trivial, because it uses Fermi liquid relations
valid at equilibrium. For the Kondo model, a perturbative approach in the
exchange constant has been used, which is valid when either $V_{ds}$ or $B$
are much larger than $T_{K}$.\cite{ros2}

Another approximation for the Anderson model is perturbation theory in $%
U/(\pi \Delta )$, where $U$ is the Coulomb repulsion and $\Delta $ is the
resonant level width.\cite{yos,horf,hor,hor3} It can be extended naturally
to the nonequilibrium case using the Keldysh formalism.\cite
{kel,lif,mahan,niu} In the equilibrium case, the second order approximation
has been used for several nanoscopic systems.\cite{lady,revi,mir,pro,pc,ogu}
A shortcoming of this method is that it cannot describe the exponential
decrease of $T_{K}$ in the Kondo regime as $U$ is increased. One way to
avoid this problem is to use renormalized perturbation theory (RPT).\cite
{he1,he2} Another method to extend the validity of the approximation to
larger values of $U$ is to use an interpolative perturbative approach (IPA),
\cite{ipa1,ipa2,levy,kaju} which corrects the second-order result in order to
reproduce exactly the atomic limit $U/\Delta \rightarrow +\infty $. The IPA
has been shown to describe well the conductance through a QD for $U\leq
8\Delta $.\cite{pro} The results agree with those obtained more recently
using the finite temperature density matrix renormalization group method.%
\cite{maru} In addition, comparison of the spin dependent IPA \cite{lady,pc}
with exact diagonalization in finite systems shows very good agreement for $%
U=6.25\Delta $.\cite{pc} Note that in some experimental situations $\Delta
\sim 0.15$ meV, $U\sim 0.6$ meV ($U/\Delta =4$),\cite{gold1,gold2} while in
recent nonequilibrium experiments $2\Delta \sim 0.33$ meV, $U\sim 1.2$ meV ($%
U/\Delta =7.3$).\cite{ama} The extension of perturbation theory in $U^{2}$
to the nonequilibrium case has been considered by Hershfield, Davies and
Wilkins.\cite{hersh} They found that for finite bias voltage $V_{ds}$, the
current is conserved only in the symmetric Anderson model, in which the
on-site energy at the dot $E_{d}=-U/2+(\mu _{L}+\mu _{R})/2$, where $\mu
_{L} $ $(\mu _{R})$ is the chemical potential at the left (right) lead.
Recent use of perturbation theory in $U/(\pi \Delta )$ for $V_{ds}\neq 0$
was limited to the symmetric case, for calculations of the current noise,%
\cite{hama2} and other properties including terms of third and fourth order
in $U$,\cite{fuj1,fuj2,hbo,hama} and in particular in presence of magnetic
field $B $,\cite{fuj2,hbo} motivated by recent experiments.\cite
{kogan,zum,ama} However, even in the symmetric case, the {\em spin} current
is not conserved by the approximation when $V_{ds}\neq 0$. In the Kondo
regime, an exact expression for the differential conductance has been
derived using RPT,\cite{hbo} but it is limited to $V_{ds}$ small in
comparison with $T_{K}$.

There is some controversy concerning the effect of $V_{ds}$ on the Kondo
peak in the spectral density for $B=0$. Calculations based on the NCA 
\cite{meirn,wing,ros1} and on the equation of motion method \cite{swir,monr}
predict a splitting of the peak. However, this method has limitations
in the Kondo regime, particularly in the particle-hole symmetric case,
for which the Green's function is indepedent of temperature.\cite{kash}
Instead, second-order
perturbation theory in $U$ predicts a fading of the peak with increasing 
$V_{ds}$ without splitting. It has been suggested that terms of order $U^{3}$
and $U^{4}$ might cause the peak to split.\cite{fuj1,fuj2} However, a recent
study with an improvement of the higher order corrections, which reproduces
correctly the equilibrium expressions, indicates that the spectral density
retains the qualitative features of the second-order result.\cite{hama}
Physically, one might expect that if the dot is hybridized with left and
right leads at chemical potentials $\mu _{L}$ and $\mu _{R}$ with matrix
elements $V_{L}$ and $V_{R}$ respectively, the effective distribution
function fluctuates between that corresponding to $\mu _{L}$ and $\mu _{R}$
(at least if $V_{L},V_{R}$ $\gtrsim $ $|\mu _{L}-\mu _{R}|$) and no split
Kondo peak at the Fermi level is expected.\cite{hama} However, as
stated above, different approximations lead to different results. The
experimental situation is also controversial: a fading
of the central peak in the differential conductance with increasing bias
voltage was reported for a carbon nanotube quantum dot,\cite{nyg} while a
splitting of the Kondo resonance in the spectral density was observed in a
three-terminal quantum ring.\cite{letu} Since the diameter of the ring is of
the order of 500 nm, it might be possible that to model the system as one
effective site connected to conducting leads is not a good approximation and
the space dependence of the energy distribution function plays an essential
role.\cite{pot}

In this work, we generalize the spin dependent IPA, based on perturbation
theory in $U$ up to second order, to the nonequilibrium case. The problem of the
conservation of the current for each spin is solved by a simple trick. The
partition of the Hamiltonian $H=H_{0}+H^{\prime }$ into an unperturbed part $%
H_{0}$ and a perturbation $H^{\prime }$ is in principle arbitrary. In the
past, for example $H_{0}$ has been chosen so that the Friedel-Langreth sum
rule \cite{lang} is satisfied at zero temperature.\cite{pro,kaju} Here we
also use the spin dependent version of this rule \cite{lady,he1} to
calculate the spectral density at equilibrium. For finite bias voltage $%
V_{ds}$, we determine $H_{0}$ asking that the current is conserved for each
spin projection. The formalism is applied to calculate the differential
conductance $dI/dV_{ds}$, spectral density and magnetization as a function
of bias voltage and magnetic field $B$ for various temperatures. Our results
are compared with experiment and previous calculations. Recent measurements 
\cite{kogan,zum,ama} of $dI/dV_{ds}$ report results which seem to disagree
with accurate calculations \cite{costi2,hof} of the spectral density at
equilibrium for $B\neq 0.$ However, Hewson {\it et al.} pointed out that
these experiments should not be interpreted in terms of the equilibrium
Green's functions.\cite{hbo} In spite of shortcomings of the selfconsistent
procedure at small temperatures, when both $V_{ds}$ and $B$ are small but
different from zero, important conclusions can be drawn from our results.

The paper is organized as follows. In Section II the model is described, the
nonequilibrium perturbation formalism is briefly reviewed, useful
expressions are derived, and the generalization of the IPA to the
nonequilibrium situation is presented. Section III contains the results of
the application of the method to magnetotransport, spectral density and
magnetization. Section IV contains a summary and a brief discussion. 
Some details of the
calculations were moved to the appendix.

\section{Model and approximations}

\subsection{Hamiltonian}

We consider a QD interacting with two conducting leads, one at the left and
one at the right, with chemical potentials $\mu _{L}$ and $\mu _{R}$
respectively, with $\mu _{L}$-$\mu _{R}=eV_{ds}$. As usual, the QD is
modelled by one effective site with one relevant orbital with an important
on-site Coulomb repulsion $U$ and an on-site energy $E_{d}$ controlled by
the gate voltage $V_{g}.$ The Hamiltonian is that of an Anderson model,
which is split into a noninteracting part $H_{0}$ and a perturbation 
$H^{\prime }$ as:

\begin{eqnarray}
H &=&H_{0}+H^{\prime },  \nonumber \\
H_{0} &=&\sum_{k\alpha \sigma }\varepsilon _{k\alpha }c_{k\alpha \sigma
}^{\dagger }c_{k\alpha \sigma }+\sum_{\sigma }\varepsilon _{eff}^{\sigma
}n_{d\sigma }+\sum_{k\alpha \sigma }(V_{k\alpha }c_{k\alpha \sigma
}^{\dagger }d_{\sigma }+\text{H.c.}),  \nonumber \\
H^{\prime } &=&(E_{d}-\varepsilon _{eff}^{\sigma })\sum_{\sigma }n_{d\sigma
}-B(n_{d\uparrow }-n_{d\downarrow })+Un_{d\uparrow }n_{d\downarrow }.
\label{h}
\end{eqnarray}
Here $\alpha =L,R$ refers to the left and right leads. The operator $%
c_{k\alpha \sigma }^{\dagger }$ creates an electron in the state with wave
vector $k$ and spin $\sigma $ at the lead $\alpha $. Similarly $d_{\sigma
}^{\dagger }$ creates an electron with spin $\sigma $ at the QD. The number
operator $n_{d\sigma }=d_{\sigma }^{\dagger }d_{\sigma }$, and $B$ is the
effect of an applied magnetic field on the on-site energy at the QD for each
spin The first term of $H_{0}$ describes the leads and the third its
hybridization with the QD. $H^{\prime }$ contains the dot on-site energy,
Zeeman splitting and Coulomb repulsion respectively. The two constants 
$\varepsilon _{eff}^{\sigma }$ are arbitrary since they cancel in 
$H_{0}+H^{\prime }$. However, in our perturbation treatment, the results
depend on them and they are determined selfconsistently as described in
subsection E.

\subsection{Nonequilibrium perturbation theory}

We use the notation of Lifshitz and Pitaevskii.\cite{lif} There are four
different Green's functions. For the electrons at the dot, they can be
chosen as

\begin{eqnarray}
G_{d\sigma }^{r}(t,t^{\prime }) &=&-i\theta (t-t^{\prime })\langle d_{\sigma
}(t)d_{\sigma }^{\dagger }(t^{\prime })+d_{\sigma }^{\dagger }(t^{\prime
})d_{\sigma }(t)\rangle ,  \nonumber \\
G_{d\sigma }^{a}(t,t^{\prime }) &=&i\theta (t^{\prime }-t)\langle d_{\sigma
}(t)d_{\sigma }^{\dagger }(t^{\prime })+d_{\sigma }^{\dagger }(t^{\prime
})d_{\sigma }(t)\rangle ,  \nonumber \\
G_{d\sigma }^{<}(t,t^{\prime }) &=&i\langle d_{\sigma }^{\dagger }(t^{\prime
})d_{\sigma }(t)\rangle ,  \nonumber \\
G_{d\sigma }^{>}(t,t^{\prime }) &=&-i\langle d_{\sigma }(t)d_{\sigma
}^{\dagger }(t^{\prime })\rangle .  \label{gdef}
\end{eqnarray}
The first two are the retarded and advanced Green's functions, already
present in the equilibrium case. Similar Green's functions can be
constructed involving conduction electrons. Since obviously $%
G^{a}-G^{r}=G^{<}-G^{>}$, only three Green's functions are independent, and
in the following we use this relation to eliminate $G^{>}$. Also, three
independent self-energy functions $\Sigma ^{r}$, $\Sigma ^{a}$, and $\Sigma
^{<}$ can de defined, which allow us to write a matrix Dyson's equation \cite
{lif}

\begin{equation}
{\bf G}={\bf g}+{\bf g\Sigma G},  \label{dyson}
\end{equation}
where

\begin{equation}
{\bf G=}\left( 
\begin{array}{ll}
0\text{ } & \text{ \qquad }G^{a} \\ 
G^{r}\text{ } & 2G^{<}+G^{r}-G^{a}
\end{array}
\right) ,
\end{equation}
${\bf g}$ has a similar expression in terms of the corresponding Green's
functions for the unperturbed Hamiltonian $H_{0}$, and

\begin{equation}
{\bf \Sigma =}\left( 
\begin{array}{ll}
\Sigma ^{r}-\Sigma ^{a}-2\Sigma ^{<} & \text{ }\Sigma ^{r} \\ 
\text{ \qquad }\Sigma ^{a} & \text{ }0
\end{array}
\right) .
\end{equation}
In general, the products in Eq. (\ref{dyson}) have to be understood as
convolutions in time and space. However, in this work, in which we consider
the stationary case and perturbation theory in $U$ up to second order,\cite
{hama} these products become just ordinary products involving the Fourier
transforms in $t-t^{\prime }$ of the Green's functions for the $d$ electrons 
$G_{d\sigma }^{\nu }(\omega )$ and the corresponding self energies $\Sigma
_{\sigma }^{\nu }(\omega )$. In addition, $G_{d\sigma }^{a}(\omega
)=(G_{d\sigma }^{r}(\omega ))^{*}$ and $\Sigma _{\sigma }^{a}(\omega
)=(\Sigma _{\sigma }^{r}(\omega ))^{*}$, where the asterisk denotes complex
conjugation. Therefore, our task reduces to find suitable approximations for 
$\Sigma _{\sigma }^{r}(\omega )$ and $\Sigma _{\sigma }^{<}(\omega )$.

After doing the matrix product in Eq. (\ref{dyson}), from the entries (1,2)
and (2,1) one obtains

\begin{equation}
\left[ G_{d\sigma }^{r}(\omega )\right] ^{-1}=\left[ g_{d\sigma }^{r}(\omega
)\right] ^{-1}-\Sigma _{\sigma }^{r}(\omega ),  \label{gr}
\end{equation}
and its complex conjugate. After some algebra and using previous equations,
the remaining equation (from the entry (2,2)) can be written in the form

\begin{equation}
G_{d\sigma }^{<}(\omega )=|G_{d\sigma }^{r}(\omega )|^{2}\left( \frac{%
g_{d\sigma }^{<}(\omega )}{|g_{d\sigma }^{r}(\omega )|^{2}}-\Sigma _{\sigma
}^{<}(\omega )\right) .  \label{gl}
\end{equation}

The noninteracting Green's functions can be obtained easily from the
equations of motion \cite{niu} using $H_{0}$:

\begin{eqnarray}
\left[ g_{d\sigma }^{r}(\omega )\right] ^{-1} &=&\omega -\varepsilon
_{eff}^{\sigma }-\sum_{\alpha }(E_{\alpha }(\omega )-i\Delta _{\alpha
}(\omega )),  \nonumber \\
g_{d\sigma }^{<}(\omega ) &=&2i|g_{d\sigma }^{r}(\omega )|^{2}\sum_{\alpha
}\Delta _{\alpha }(\omega )f(\omega -\mu _{\alpha }),  \label{g0}
\end{eqnarray}
where $f(\omega )$ is the Fermi function and $E_{\alpha }$, $\Delta _{\alpha
}$ for leads $\alpha =L,R$, are the real and imaginary parts of the sum

\begin{equation}
\sum_{k}\frac{|V_{k\alpha }|^{2}}{\omega -\varepsilon _{k\alpha }+i\eta }%
=E_{\alpha }(\omega )-i\Delta _{\alpha }(\omega ),  \label{delta}
\end{equation}
with $\eta $ a positive infinitesimal.

We approximate the retarded self energy as

\begin{equation}
\Sigma _{\sigma }^{r}(\omega )=E_{d}^{\sigma }-\varepsilon _{eff}^{\sigma
}+U\langle n_{d\overline{\sigma }}\rangle +\Sigma _{\sigma \text{int}%
}^{r2}(\omega ),  \label{sr}
\end{equation}
where $E_{d}^{\uparrow }=E_{d}-B$, $E_{d}^{\downarrow }=E_{d}+B$, and $%
\overline{\sigma }$ means $-\sigma $. The last term is an interpolative
expression based on the correction of order $U^{2}$, as described in
subsection D. The remaining terms would correspond to the first order
correction in $H^{\prime }$ if $\langle n_{d\overline{\sigma }}\rangle $
were evaluated with $H_{0}$. However, $\langle n_{d\overline{\sigma }%
}\rangle $ is evaluated selfconsistently using

\begin{equation}
\langle n_{d\sigma }\rangle =\frac{-i}{2\pi }\int d\omega G_{d\sigma
}^{<}(\omega ),  \label{nds}
\end{equation}
and this is equivalent to a partial summation of an infinite series of
diagrams.\cite{fuj2}

The correction for $\Sigma _{\sigma }^{<}(\omega )$ of first order in $%
H^{\prime }$ vanishes. We approximate $\Sigma _{\sigma }^{<}(\omega )$ by an
expression $\Sigma _{\sigma \text{int}}^{<}(\omega )$ based on the second
order term $\Sigma _{\sigma }^{2<}(\omega )$, also described in subsection D.

The diagrams for the corrections to the self energy of second order in $U$
are drawn in Ref. \onlinecite{lif}. In terms of the independent unperturbed
Green's functions the expressions are

\begin{eqnarray}
\Sigma _{\uparrow }^{r2}(\omega ) &=&U^{2}\int \frac{d\omega _{1}}{2\pi }%
\int \frac{d\omega _{2}}{2\pi }  \nonumber \\
&&[g_{\uparrow }^{r}(\omega _{1})g_{\downarrow }^{r}(\omega
_{2})g_{\downarrow }^{<}(\omega _{1}+\omega _{2}-\omega )  \nonumber \\
&&+g_{\uparrow }^{r}(\omega _{1})g_{\downarrow }^{<}(\omega
_{2})g_{\downarrow }^{<}(\omega _{1}+\omega _{2}-\omega )  \nonumber \\
&&+g_{\uparrow }^{<}(\omega _{1})g_{\downarrow }^{r}(\omega
_{2})g_{\downarrow }^{<}(\omega _{1}+\omega _{2}-\omega )  \nonumber \\
&&+g_{\uparrow }^{<}(\omega _{1})g_{\downarrow }^{<}(\omega
_{2})g_{\downarrow }^{a}(\omega _{1}+\omega _{2}-\omega )],  \label{sr2}
\end{eqnarray}

\begin{eqnarray}
\Sigma _{\uparrow }^{<2}(\omega ) &=&-U^{2}\int \frac{d\omega _{1}}{2\pi }%
\int \frac{d\omega _{2}}{2\pi }  \nonumber \\
&&g_{\uparrow }^{<}(\omega _{1})g_{\downarrow }^{<}(\omega
_{2})g_{\downarrow }^{>}(\omega _{1}+\omega _{2}-\omega ),  \label{sl2}
\end{eqnarray}
and the same interchanging spin up and down, with $g_{\sigma }^{>}(\omega
)=g_{\sigma }^{<}(\omega )+2%
\mathop{\rm Im}%
g_{\sigma }^{r}(\omega )$. It can be shown that these expressions are
equivalent to those given by Hershfield {\it et al}.\cite{hersh} These
integrals which have the form of convolutions in frequency can be calculated
conveniently using fast Fourier transform to the time representation.\cite
{hor3,czy} For the case of constant $E_{\alpha }(\omega )$ and $\Delta
_{\alpha }(\omega )$, it is more convenient to evaluate one of the integrals
analytically, as shown in the appendix.

Although it is not convenient for the numerical evaluation, it is
interesting to express the above self-energy corrections in terms of the
unperturbed spectral density of states at the dot $\rho _{d\sigma
}^{0}(\omega )=-%
\mathop{\rm Im}%
g_{d\sigma }^{r}(\omega )/\pi $. In terms of this density

\begin{eqnarray}
g_{d\sigma }^{r}(\omega ) &=&\int d\epsilon \frac{\rho _{d\sigma
}^{0}(\epsilon )}{\omega -\epsilon +i\eta },  \nonumber \\
g_{d\sigma }^{<}(\omega ) &=&2i\pi \rho _{d\sigma }^{0}(\omega )\tilde{f}%
(\omega ),  \label{gro}
\end{eqnarray}
where $\tilde{f}(\omega )$ is a weighted distribution function of the Fermi
functions at the two leads:

\begin{equation}
\tilde{f}(\omega )=\frac{\sum_{\alpha }\Delta _{\alpha }(\omega )f(\omega
-\mu _{\alpha })}{\sum_{\alpha }\Delta _{\alpha }(\omega )}.  \label{df}
\end{equation}
Replacing Eqs. (\ref{gro}) in Eqs. (\ref{sr2}) and (\ref{sl2}) we obtain
after some algebra

\begin{eqnarray}
\Sigma _{\uparrow }^{r2}(\omega ) &=&U^{2}\int d\epsilon _{1}d\epsilon
_{2}d\epsilon _{3}\frac{\rho _{d\uparrow }^{0}(\epsilon _{1})\rho
_{d\downarrow }^{0}(\epsilon _{2})\rho _{d\downarrow }^{0}(\epsilon _{3})}{%
\omega +\epsilon _{3}-\epsilon _{1}-\epsilon _{2}+i\eta }  \nonumber \\
&&\times [(1-\tilde{f}(\epsilon _{1}))(1-\tilde{f}(\epsilon _{2}))\tilde{f}%
(\epsilon _{3})  \nonumber \\
&&+\tilde{f}(\epsilon _{1})\tilde{f}(\epsilon _{2})(1-\tilde{f}(\epsilon
_{3}))],  \label{srro}
\end{eqnarray}

\begin{eqnarray}
\Sigma _{\uparrow }^{<2}(\omega ) &=&-2i\pi U^{2}\int d\epsilon
_{1}d\epsilon _{2}\rho _{d\uparrow }^{0}(\epsilon _{1})\rho _{d\downarrow
}^{0}(\epsilon _{2})\rho _{d\downarrow }^{0}(\epsilon _{1}+\epsilon
_{2}-\omega )  \nonumber \\
&&\times \tilde{f}(\epsilon _{1})\tilde{f}(\epsilon _{2})(1-\tilde{f}%
(\epsilon _{1}+\epsilon _{2}-\omega )).  \label{slro}
\end{eqnarray}
At equilibrium $\mu _{L}=\mu _{R}$. Then $\tilde{f}(\omega )$ becomes the
usual distribution function and $\Sigma _{\sigma }^{r2}(\omega )$ coincides
with the result of ordinary perturbation theory. Eqs. (\ref{srro}) and (\ref
{slro}) are used in subsection D to construct the corresponding
interpolating expressions $\Sigma _{\sigma \text{int}}^{r2}(\omega )$ and $%
\Sigma _{\sigma \text{int}}^{<}(\omega )$.

\subsection{The current}

Following Meir and Wingreen,\cite{meir} the current with spin $\sigma $
flowing between the left lead and the dot is

\begin{equation}
j_{L\sigma }=\frac{2ie}{h}\int d\omega \Delta _{L}(\omega )\left[ 2if(\omega
-\mu _{L})%
\mathop{\rm Im}%
G_{d\sigma }^{r}(\omega )+G_{d\sigma }^{<}(\omega )\right] \text{.}
\label{jl}
\end{equation}
Similarly, the current with spin $\sigma $ flowing between the dot and the
right lead is

\begin{equation}
j_{R\sigma }=-\frac{2ie}{h}\int d\omega \Delta _{R}(\omega )\left[
2if(\omega -\mu _{R})%
\mathop{\rm Im}%
G_{d\sigma }^{r}(\omega )+G_{d\sigma }^{<}(\omega )\right] \text{.}
\label{jr}
\end{equation}
Of course, since the current is conserved one should have

\begin{equation}
j_{L\sigma }=j_{R\sigma },  \label{je}
\end{equation}
but perturbation theory does not satisfy this equality for generic values of
the parameters.\cite{hersh} In this work we determine $\varepsilon
_{eff}^{\sigma }$ imposing Eqs. (\ref{je}).

\subsection{The interpolative perturbative approach (IPA)}

Extending previous ideas,\cite{ipa1,ipa2,levy,kaju} we want to replace the
second-order contributions to the self energies $\Sigma _{\sigma
}^{r2}(\omega )$ and $\Sigma _{\sigma }^{<2}(\omega )$, by other ones $%
\Sigma _{\sigma \text{int}}^{r2}(\omega )$ and $\Sigma _{\sigma \text{int}%
}^{<}(\omega )$, which coincide with the previous ones to order $U^{2}$ for
small $U$, but also reproduce the high frequency and atomic limits. For the
sake of clarity in the exposition we choose $\sigma =\uparrow $. The final
results are also valid interchanging spin up and down.

We propose \cite{kaju}

\begin{equation}
\Sigma _{\uparrow \text{int}}^{r2}(\omega )=\frac{A_1\Sigma _{\uparrow
}^{r2}(\omega )}{1-A_2\Sigma _{\uparrow }^{r2}(\omega )},  \label{s2int}
\end{equation}
where $A_1$ is determined so that $\Sigma _{\uparrow \text{int}}^{r2}(\omega )$
reproduces the leading behavior at high frequencies, and afterwards $A_2$ is
determined to reproduce the exact result in the atomic limit $\Delta
_{\alpha }(\omega )/U\rightarrow 0$.

Up to order $1/\omega $, the exact self energy is determined by the first
and second moments of the spectral density of $d$ states $\rho _{d\sigma
}(\omega )$, which can be evaluated independently of $\rho _{d\sigma
}(\omega )$ using particular commutators.\cite{nolt} Proceeding as in the
equilibrium case,\cite{kaju} allowing dependence on spin,\cite{lady,pc} and
using Eqs. (\ref{gr}) and (\ref{sr}), one obtains that the correct leading
behavior of the retarded second-order self energy is

\begin{equation}
\Sigma _{\uparrow \text{int}}^{r2}(\omega )=\langle n_{d\downarrow }\rangle
(1-\langle n_{d\downarrow }\rangle )U^{2}\omega ^{-1}+O(\omega ^{-2}).
\end{equation}
For $\omega \rightarrow \infty $, the integrals over $\epsilon _{i}$ in Eq. (%
\ref{srro}) decouple and then

\begin{equation}
\Sigma _{\uparrow }^{r2}(\omega )=\langle n_{d\downarrow }^{0}\rangle
(1-\langle n_{d\downarrow }^{0}\rangle )U^{2}\omega ^{-1}+O(\omega ^{-2}),
\end{equation}
where $\langle n_{d\sigma }^{0}\rangle =(-i/2\pi )\int d\omega g_{d\sigma
}^{<}(\omega )$ is the expectation value of $n_{d\sigma }$ calculated using
the noninteracting lesser Green's function [see Eq. (\ref{gro})]. From the
above equations, one sees that choosing

\begin{equation}
A_1=\langle n_{d\downarrow }\rangle (1-\langle n_{d\downarrow }\rangle
)/[\langle n_{d\downarrow }^{0}\rangle (1-\langle n_{d\downarrow
}^{0}\rangle ],  \label{a}
\end{equation}
the moments of $\rho _{d\sigma }(\omega )$ up to the second one are
reproduced exactly.

In the atomic limit, it can be easily verified using equations of motion,%
\cite{niu} that the exact retarded Green's function is

\begin{equation}
G_{d\uparrow }^{r,\text{at}}(\omega )=\frac{1-\langle n_{d\downarrow
}\rangle }{\omega -E_{d}^{\uparrow }+i\eta }+\frac{\langle n_{d\downarrow
}\rangle }{\omega -E_{d}^{\uparrow }-U+i\eta }.  \label{gat}
\end{equation}
From Eqs. (\ref{gr}) and (\ref{sr}), this implies

\begin{equation}
\Sigma _{\uparrow \text{int}}^{r2,\text{at}}(\omega )=\frac{\langle
n_{d\downarrow }\rangle (1-\langle n_{d\downarrow }\rangle )U^{2}}{\omega
-E_{d}^{\uparrow }-(1-\langle n_{d\downarrow }\rangle )U+i\eta }.
\label{sat}
\end{equation}
The usual contribution in $U^{2}$ to the retarded self energy can be
calculated easily in the atomic limit using Eq. (\ref{srro}), because the
noninteracting spectral densities $\rho _{d\sigma }^{0}(\omega )$ become
delta functions. The result can be written in the form

\begin{equation}
\Sigma _{\uparrow }^{r2,\text{at}}(\omega )=\frac{\langle n_{d\downarrow
}^{0}\rangle (1-\langle n_{d\downarrow }^{0}\rangle )U^{2}}{\omega
-\varepsilon _{eff}^{\uparrow }+i\eta }.  \label{sat2}
\end{equation}
From Eqs. (\ref{s2int}), (\ref{a}), (\ref{sat}) and (\ref{sat2}), the
coefficient $A_2$ is determined. The final result is

\begin{equation}
\Sigma _{\uparrow \text{int}}^{r2}(\omega )=\frac{\langle n_{d\downarrow
}\rangle (1-\langle n_{d\downarrow }\rangle )\Sigma _{\uparrow }^{r2}(\omega
)}{\langle n_{d\downarrow }^{0}\rangle (1-\langle n_{d\downarrow
}^{0}\rangle )+[\varepsilon _{eff}^{\uparrow }-E_{d}^{\uparrow }-(1-\langle
n_{d\downarrow }\rangle )U](\Sigma _{\uparrow }^{r2}(\omega )/U^{2})}.
\label{intr}
\end{equation}
This result has the same form as in the spin dependent situation in
equilibrium.\cite{lady,pc} However, the expectation values are calculated
with distribution functions out of equilibrium.

From the equations of motion in the atomic limit (see section 3 of Ref. \cite
{niu}{)}, one obtains that the lesser Green's function for $V_{k\alpha }=0$
is the sum of two delta functions $G_{d\uparrow }^{<,\text{at}}(\omega
)=ia\delta (\omega -E_{d}^{\uparrow })+ib\delta (\omega -E_{d}^{\uparrow
}-U) $, where $a$ and $b$ are two real unknown constants. Using Eq. (\ref
{gat}), we can write $G_{d\uparrow }^{<,\text{at}}(\omega )=-2iy(\omega )%
\mathop{\rm Im}%
G_{d\uparrow }^{r,\text{at}}(\omega )$, where the function $y(\omega )$ is
unknown. Electron-hole symmetry in the symmetric Anderson model imposes that 
$y(-\omega )=1-y(\omega )$. At equilibrium, $\mu _{L}=\mu _{R}=\mu $ and it
can be shown from their definitions [Eqs. (\ref{gdef})]\cite{meir} that $%
y(\omega )=f(\omega -\mu )$. Out of equilibrium, $y(\omega )$ is ill defined
at $V_{k\alpha }=0$, since the dot is disconnected to the leads, while even
for small non-zero $V_{k\alpha }$ the problem is not trivial. In the general
case, we assume that $y(\omega )=\tilde{f}(\omega )$, the average
distribution function defined in Eq.(\ref{df}). This has the correct limit
at equilibrium or when the limit $V_{k\alpha }\rightarrow 0$ is taken for
one of the leads first and then for the other. Using this choice, Eqs. (\ref
{gr}), (\ref{gl}) and (\ref{g0}) imply for the lesser self energy in the
atomic limit

\begin{equation}
\Sigma _{\uparrow \text{int}}^{<,\text{at}}(\omega )=2i\tilde{f}(\omega )%
\mathop{\rm Im}%
\Sigma _{\uparrow \text{int}}^{r2,\text{at}}(\omega ).  \label{slat}
\end{equation}
On the other hand, using Eqs. (\ref{g0}), (\ref{slro}) and (\ref{sat2}), we
can write for the second order correction in the same limit

\begin{equation}
\Sigma _{\uparrow }^{2<,\text{at}}(\omega )=2i\tilde{f}(\omega )%
\mathop{\rm Im}%
\Sigma _{\uparrow }^{r2,\text{at}}(\omega ).  \label{slat2}
\end{equation}
From these equations, and taking into account that $\Sigma _{\uparrow \text{%
int}}^{r2}(\omega )$ coincides with $\Sigma _{\uparrow }^{r2}(\omega )$ for
small $U$ and with $\Sigma _{\uparrow }^{r2,\text{at}}(\omega )$, the
simplest interpolative expression for $\Sigma _{\uparrow }^{<}(\omega )$ is

\begin{equation}
\Sigma _{\uparrow \text{int}}^{<}(\omega )=\frac{%
\mathop{\rm Im}%
\Sigma _{\uparrow \text{int}}^{r2}(\omega )}{%
\mathop{\rm Im}%
\Sigma _{\uparrow }^{r2}(\omega )}\Sigma _{\uparrow }^{2<}(\omega )\text{, }
\label{intl}
\end{equation}
when $%
\mathop{\rm Im}%
\Sigma _{\uparrow }^{r2}(\omega )\neq 0$, and zero otherwise. Note that if $%
\mathop{\rm Im}%
\Sigma _{\uparrow }^{r2}(\omega )=0$, also $\Sigma _{\uparrow }^{2<}(\omega
) $ $=0$, as can be seen from Eqs. (\ref{srro}) and (\ref{slro}).

\subsection{The selfconsistency problem}

Eq. (\ref{intr}) replaced in Eq. (\ref{sr}) and the same interchanging $%
\uparrow $ and $\downarrow $ define the retarded self energies. Similarly
Eq. (\ref{intl}), and the same for spin down are the lesser self energies
used. The expressions for $\Sigma _{\sigma }^{r2}(\omega )$ and $\Sigma
_{\sigma }^{2<}(\omega )$ for constant $\Delta _{\alpha }$ are given in the
appendix. The self energies replaced in Eqs. (\ref{gr}) to (\ref{g0}) give
the Green's functions for the interacting system that depend in general on
four unknown quantities: $\varepsilon _{eff}^{\sigma }$ and $\langle
n_{d\sigma }\rangle $. They are determined by the selfconsistent solution of
Eqs. (\ref{nds}) and (\ref{je}). The resulting values of 
$\varepsilon _{eff}^{\sigma }$ do not have any particular physical meaning. 

In the present work, we present results for the symmetric Anderson model, $%
E_{d}=-U/2$, $\mu _{L}=-$ $\mu _{R}=eV_{ds}/2$, $\Delta _{L}=\Delta _{R}$
independent of energy, in presence of a magnetic field $B$. In this case,
due to electron-hole symmetry $\varepsilon _{eff}^{\sigma }=-\varepsilon
_{eff}^{\bar{\sigma}}$ and $j_{L\sigma }=j_{R\bar{\sigma}}$. This reduces
the problem to two selfconsistent equations with two unknowns.

Unfortunately, for small temperature $T$, small non-zero $V_{ds}$ and small
non-zero $B$, there were instabilitities in the numerical algorithm used.
Also in some regions no solution could be found, while in other cases the
solution jumped to another value. Therefore our results should be regarded
as complementary to those obtained using renormalized perturbation theory
(RPT), which correspond to $T=0$ and small $V_{ds}$.\cite{hbo} We have tried
an alternative approach, relaxing the condition of conservation of current
and fixing $\varepsilon _{eff}^{\sigma }$ either at the value that satisfies
the Friedel-Langreth sum rule (see below) or $\varepsilon _{eff}^{\sigma }=0$%
. In the former case, the solution of Eqs. (\ref{nds}) for $\langle
n_{d\sigma }\rangle $ was lost for very small $V_{ds}$, while in the second,
the peak in the differential conductance $dI/dV_{ds}$ was already split for
very small values of $B$ (of the order of $T_{K}/30$) in contradiction with
the results of RPT. Therefore these alternatives were abandoned.

In general, we have started from the solution for large $V_{ds}$ (with small
reasonable $|\varepsilon _{eff}^{\sigma }|)$, which was easy to find, and
used it as an initial guess for lower $V_{ds}$. The procedure was repeated
until reaching a small non-zero value of $V_{ds}$ or some instability.

For $V_{ds}=0$, the problem is ill posed since the current is conserved for
any $\varepsilon _{eff}^{\sigma }$. Also the first derivative of $j_{L\sigma
}-j_{R\sigma }$ with respect to $V_{ds}$ vanishes and the second is very
small for any $\varepsilon _{eff}^{\sigma }$. For $V_{ds}=T=0$, we have
calculated the spectral density replacing the condition $j_{L\sigma
}=j_{R\sigma }$ by the Friedel-Langreth sum rule \cite{lang} for non-zero
magnetic field, which for constant $\Delta =\Delta _{R}(\omega )+\Delta
_{L}(\omega )$ can be written in the form \cite{lady,he1}

\begin{equation}
\pi \Delta \rho _{d\sigma }(0)=\frac{\Delta ^{2}}{\Delta ^{2}+\left[
\varepsilon _{eff}^{\sigma }+\Sigma _{\sigma }^{r}(0)\right] ^{2}}=\sin
^{2}(\pi \langle n_{d\sigma }\rangle ).  \label{fri}
\end{equation}

\section{Numerical results}

In this section we take $\Delta _{R}(\omega )=\Delta _{L}(\omega )$
independent of energy and $\Delta =\Delta _{R}+\Delta _{L}=1$ as the unit of
energy. The origin of energy ($\omega=0$) is set at $(\mu _{L}+\mu _{R})/2$.
We have chosen $U=5$ for the numerical analysis because we want $U$
large enough so that the system is at the Kondo regime, but for $U>6.5$, the
magnetic susceptibility for $V_{ds}=T=0$ decreases with increasing $U$,
indicating the failure of the IPA for a quantitative analysis at small
non-zero magnetic field. These parameters lead to a Kondo temperature
$T_{K}=0.45$ as determined from the width of the spectral density.

To represent the spectral density for $V_{ds}=T=0$,
we have determined selfconsistently the occupation numbers $\langle
n_{d\sigma }\rangle $ and effective energies $\varepsilon _{eff}^{\sigma }$
using the Friedel-Langreth sum rule, while to calculate the conductivity at
finite $V_{ds}$ we have replaced this rule by the condition of conservation
of current as described above.

In Fig. 1 we represent the differential conductance $dI/dV_{ds}$ as a
function of the applied voltage between drain and source $V_{ds}$ for zero
temperature, and compare it with the total spectral density of states $\rho
_{d}(\omega )=\rho _{d\uparrow }(\omega )+\rho _{d\downarrow }(\omega )$ for 
$V_{ds}=0$ and $\omega =eV_{ds}$. The spectral density at $B=0$ shows a
central peak corresponding to the Kondo peak with a half width at half
maximum $T_{K}=0.45$ and two shoulders at $\omega =E_{d}=-2.5$ and $\omega
=E_{d}+U=2.5$, which tend to separate with increasing magnetic field. The
central peak shows a splitting which is tiny for $B=0.2$ (not shown) and
increases with applied magnetic field. The splitting of the peak is larger
than the Zeeman splitting $2B$ by a factor $\sim 1.5$ for $B=0.3$ and
increases to slightly below 2 ($\sim 1.8$) for $B=1.5$. This behavior agrees
with previous studies of the spectral density under an applied magnetic
field for large enough $U$ \cite{costi,meirn,moore,logan2,he2,fuj2}

The differential conductance $dI/dV_{ds}$ was obtained by numerical
differentiation of the total current $I=j_{L\uparrow }+j_{L\downarrow
}=j_{R\uparrow }+j_{R\downarrow }$. For $0<B<0.5$ and $0<V_{ds}<1.5$, we
could not find a self consistent solution of the system of equations, and
therefore we are not able to describe how the peak in the differential
conductance splits with applied magnetic field at $T=0$. As described below,
the situation improves with applied temperature. For $B=0$, the peak in $%
dI/dV_{ds}$ is narrower than that of the spectral density with a half width
at half maximum of 0.383. In addition the shoulders are displaced from the
central peak. In general, the split peaks in $dI/dV_{ds}$ are displaced to
higher $|eV_{ds}|$ when compared with the same peaks in $\rho _{d}(\omega )$%
. Because of the lack of results for small non vanishing magnetic field and
voltage, our calculations for $T=0$ are not enough to establish if this
tendency persists for small splittings.

In Fig. 2, we show the evolution of the spectral density with applied
voltage. For $B=0$, the Kondo peak decreases slowly without splitting. For $%
V_{ds}=3$ (not shown), $\rho _{d}(\omega )$ is flat near $\omega =0$.
Further increase in $V_{ds}$ leads to only very small changes in $\rho
_{d}(\omega )$. For large magnetic field, the changes in $\rho _{d}(\omega )$
with $V_{ds}$ are less dramatic. For large $V_{ds}$ the spectral density for
different applied magnetic fields looks similar, but somewhat flatter and
broader for larger $B$.

In Fig. 3 we show the magnetization $m=\langle n_{d\uparrow }\rangle
-\langle n_{d\downarrow }\rangle $ as a function of applied voltage between
drain and source for different values of the magnetic field and temperature.
Because of our problems with the self-consistent solution for $T=0$, the
values of $B>T_{K}=0.45$ taken at zero temperature are large enough so that
the Kondo effect is at least partially destroyed. In this case, an increase
in $V_{ds}$ leads to a decrease in the magnetization which is expected as a
consequence of the flattening of the spectral density of states discussed
above. These curves for $T=0$ and comparatively large values of $B$ show an
initial slow decrease with increasing applied voltage, an inflection point
for $eV_{ds}\sim 4B$, and an asymptotic approach to $m=0$ for large $V_{ds}$%
. Instead for small $B$ (see the curve for $B=0.2$ and $T=3$), the
magnetization first increases and then decreases with applied voltage. This
can be interpreted in terms of a destruction of the Kondo singlet state for $%
V_{ds}\lesssim 2T_{K}$, which moves the system towards the local moment
regime, followed by the same effects of flattening of the spectral density
present for higher $B$.

For temperatures $T\sim T_{K}/2$, we have obtained some results in the
problematic region for selfconsistency ($0<B<T_{K}$ and $0<V_{ds}\lesssim
2T_{K}$). Some of them are shown in Fig. 4. However, still the value of $B$
at which the splitting of the central peak takes place could not be
identified precisely and abrupt changes in the solution take place for
example for $B=T=0.3$ and $B=T=0.2$. In spite of this, these results and
those for $T=0.5$ shown below, suggest a transition from one maxima to two
maxima in the differential conductance $dI/dV_{ds}$ as a function of applied
voltage for $B_{c}\sim 0.3=(2/3)T_{K}$. This value is consistent with
that obtained by Hewson and coworkers \cite{hbo} $B_{c}=0.584T_{K}$.
Comparison of all split maxima of $dI/dV_{ds}$ with the corresponding ones
of the spectral density, shows again that all the former lie at values of 
$|eV_{ds}|$ larger than the corresponding values of $|\omega |$ for the
maxima of $\rho _{d}(\omega )$. This provides a natural explanation of the
disagreement of the splitting of the peaks observed in experiments of the
differential conductance in presence of a magnetic field, \cite
{kogan,zum,ama} when compared with accurate results of the spectral density 
\cite{costi2}. The observed splitting for sufficiently large field was found
to be higher than the corresponding one of the spectral density.

In Fig. 5, we show the evolution of $\rho _{d}(\omega )$ with applied
voltage for $T=0.3$. For $B=0.2$ and small $V_{ds}$, the spectral density
displays only one maximum at $\omega =0$, although its components $\rho
_{d\uparrow }(\omega )$ and $\rho _{d\downarrow }(\omega )$ have maxima at $%
\omega =-\omega _{M}$ and $\omega =\omega _{M}$ respectively, with $\omega
_{M}>0$, as shown in Fig. 5 (a) for $eV_{ds}=0.1$. We remind the reader that
due to our choice of parameters $\rho _{d\downarrow }(\omega )=\rho
_{d\uparrow }(-\omega )$. As in the case of zero temperature (Fig. 2), the
changes driven by the applied voltage on $\rho _{d\sigma }(\omega )$ are
smaller for higher magnetic fields.

In Fig. 6 we show $dI/dV_{ds}$ as a function of $V_{ds}$ for $T=0.5$, and
the spectral density $\rho _{d}(\omega )$ for $eV_{ds}=0.1$. As before and
to facilitate comparison, the abscissa of $\rho _{d}(\omega )$ is chosen in
such a way that $\omega $ in Fig. 6 (a) is the same as $eV_{ds}$ in Fig. 6
(b). The scales are the same as in Fig. 4. For this temperature, there were
no instabilities in the algorithm for selfconsistent equations and we could
follow the splitting of the central Kondo peak. The transition from one
maxima to two in the differential conductance takes place slightly below $%
B=0.3$. With increasing temperature, both the spectral density and the
differential conductance become flatter, and the changes are more noticeable
for small vales of the magnetic field. Again, the structure of $dI/dV_{ds}$
is broader than that of $\rho _{d}(\omega )$ and the maxima are more
separated from $\omega =0$.

\section{Summary and discussion}

We have generalized the spin dependent interpolative perturbative
approximation (IPA), based on second-order perturbation theory in the
Coulomb repulsion $U$, to the nonequilibrium case. This permits to extend
the validity of the results of nonequilibrium second order perturbation
theory in $U/(\pi \Delta )$ to higher values of $U$. The effective
unperturbed energy at the dot has been determined selfconsistently asking
that the current for each spin is conserved. This allows us to correct the
shortcoming of ordinary perturbation theory in $U/(\pi \Delta )$ that the
spin current is not conserved, except for the symmetric Anderson model
without an applied magnetic field. 

We have applied the method to study the effects of a magnetic field $B$ on
the differential conductance $dI/dV_{ds}$, starting from the symmetric
Anderson model in which the on-site energy $E_{d}=-U/2$, setting the origin
of energies at the average of the chemical potentials, and the resonant
level width for both leads are equal $(\Delta _{L}=\Delta _{R})$ and
independent of energy. The selfconsistent procedure fails for temperatures
below the Kondo temperature $T_{K}$, when $0<B<T_{K}$ and $0<V_{ds}\lesssim
2T_{K}.$ 

The procedure can in principle be applied to the general case (without
starting from the symmetric case). Then, one has to solve four self consistent
equations for the effective occupations up and down, and the corrresponding
effective unperturbed levels. Similar problems were solved for the equilibrium
case in presence of a magnetic field \cite{lady} and to calculate persistent
currennts in a ring with an odd number $N$ of electrons.\cite{pc} In the
latter case however, the procedure failed for large $N$. In the present case,
one might expect more numerical difficulties after our experience with the simpler
case treated here.

In spite of the above mentioned failure, our results indicate that the critical
magnetic field $B_{c}$ for the splitting of the Kondo resonance in 
$dI/dV_{ds}$ as a function of $V_{ds}$ is near and slightly below $(2/3)T_{K}$. 
This is consistent with the results of renormalized perturbation theory 
\cite{hbo} which give $B_{c}=0.584T_{K}$. 

 In the region of split peaks, the
splitting is larger than the corresponding one of the spectral density of
states $\rho _{d}(\omega )$ as a function of $\omega$. This provides a
natural explanation of why recent experiments of the differential
conductance in presence of a large magnetic field give a splitting larger
than two times the Zeeman splitting, as expected for the spectral density of
states of states at equilibrium.\cite{kogan,zum,ama} Instead, previous results 
using slave bosons obtained a reduction of the peak splitting.\cite{dong}   

The effect of increasing bias voltage on the spectral density $\rho
_{d}(\omega )$ is just an overall flattening of the energy dependence,
without introducing new peaks. This is in agreement with recent calculations
in the symmetric case \cite{hama} including terms of third and fourth order
in $U$. 

For small applied magnetic field, so that the system is in the Kondo regime,
the magnetization as a function of the applied bias voltage $V_{ds}$ first
increases because due to the destruction of the Kondo singlet state, the
system enters in the localized moment regime. For $V_{ds}\gtrsim 2T_{K}$, or 
$B>T_{K}$, the effect of flattening of the spectral density dominates and
the magnetization decreases with increasing $V_{ds}$.

\section*{Acknowledgments}

I thank M.J. S\'{a}nchez for many helpful discussions. I am partially
supported by CONICET. This work was sponsored by PICT 03-13829 of ANPCyT.

\appendix

\section{Self energies for constant $\Delta _{\alpha }$}

Usually for problems in QD's, the density of states of the wide band of
conduction electrons at the leads, as well as the hybridization with the dot 
$V_{k\alpha }$ can be taken as constant within the narrow energy range of $%
T_{K}$ or other energy scales in the problem.\cite{pro} In this case $%
E_{\alpha }$ and $\Delta _{\alpha }$ become independent of energy and one of
the integrals in Eqs. (\ref{sr2}) and (\ref{sl2}) can be done analytically. $%
E_{\alpha }$ merely renormalizes $E_{d}$.

Decomposing $g_{d\sigma }^{<}(\omega )$ as well as products of unperturbed
Green functions as sums of simple fractions with single poles times Fermi
functions, and decomposing products of Fermi functions using

\begin{equation}
f(x)f(y)=\frac{1}{\exp \left[ (x+y)/T\right] -1}(f(-x)-f(y)),  \label{ff}
\end{equation}
the first integral in frequency is decomposed into a sum of terms of the form

\begin{equation}
I_{0}=\int d\omega f(\omega )\frac{1}{\omega -a\pm ib},  \label{i0}
\end{equation}
with $a$ and $b$ real. This integral can be evaluated using the digamma
function:\cite{abra}

\begin{equation}
I_{0}-\int d\omega f(\omega )\frac{1}{\omega \pm i\eta }=\psi (\frac{1}{2}+%
\frac{b\pm ia}{2\pi T})-\psi (\frac{1}{2}).  \label{dig}
\end{equation}

After a lengthy algebra, the corrections of order $U^{2}$ to the self
energies become

\begin{eqnarray}
\Sigma _{\uparrow }^{r2}(\omega ) &=&\left( \frac{U}{2\pi }\right) ^{2}\int
d\nu [g_{d\uparrow }^{r}(\omega -\nu )I_{1}(-\nu )  \nonumber \\
&&+g_{d\uparrow }^{<}(\omega -\nu )I_{2}(\nu )],  \label{s2a} \\
\Sigma _{\uparrow }^{<2}(\omega ) &=&-\left( \frac{U}{2\pi }\right) ^{2}\int
d\nu g_{d\uparrow }^{<}(\omega -\nu )I_{1}(\nu ),  \nonumber
\end{eqnarray}
and the same interchanging spins up and down, with

\begin{eqnarray}
I_{1}(\nu ) &=&\frac{2}{\Delta ^{2}}%
\mathop{\rm Re}%
\{\frac{2i\Delta }{\nu (\nu +2i\Delta )}\sum_{\alpha }[(A_{\alpha }+B_{\bar{%
\alpha}})(\varphi _{2\alpha }^{*}(-\nu )-\varphi _{1\alpha })  \label{i12} \\
&&+(A_{\alpha }+B_{\alpha })(\varphi _{2\alpha }(\nu )-\varphi _{1\alpha
}^{*})]\},  \nonumber \\
I_{2}(\nu ) &=&-\sum_{\alpha }\frac{2i\Delta _{\alpha }}{\nu (\nu +2i\Delta )%
}[\varphi _{1\alpha }+\varphi _{1\alpha }^{*}-\varphi _{2\alpha }(\nu
)-\varphi _{2\alpha }^{*}(-\nu )],  \nonumber
\end{eqnarray}
where $\Delta =\Delta _{L}+\Delta _{R}$, $\bar{\alpha}=R$ $(L)$ if $\alpha
=L $ $(R)$, and $\varphi _{1\alpha }$ and the functions $\varphi _{2\alpha
}(\nu )$, $A_{\alpha }(\nu )$ and $B_{\alpha }(\nu )$ are

\begin{eqnarray}
\varphi _{1\alpha } &=&\psi (\frac{1}{2}+\frac{\Delta +i(\varepsilon
_{eff}^{\downarrow }-\mu _{\alpha })}{2\pi T}),  \nonumber \\
\varphi _{2\alpha }(\nu ) &=&\psi (\frac{1}{2}+\frac{\Delta +i(\varepsilon
_{eff}^{\downarrow }-\mu _{\alpha }-\nu )}{2\pi T}),  \nonumber \\
A_{\alpha }(\nu ) &=&\frac{\Delta _{\alpha }^{2}}{\exp (\nu /T)-1}, 
\nonumber \\
B_{\alpha }(\nu ) &=&\frac{\Delta _{L}\Delta _{R}}{\exp [(\nu +\mu _{\alpha
}-\mu _{\bar{\alpha}})/T]-1}.  \label{last}
\end{eqnarray}
In equilibrium, when $\mu _{L}=\mu _{R}$, the expression for $\Sigma
_{\uparrow }^{r2}(\omega )$ can be easily shown to coincide with that given
by Horvati\'{c} and Zlati\'{c}.\cite{horf}

\medskip


\begin{figure}
\hskip1.0cm\psfig{file=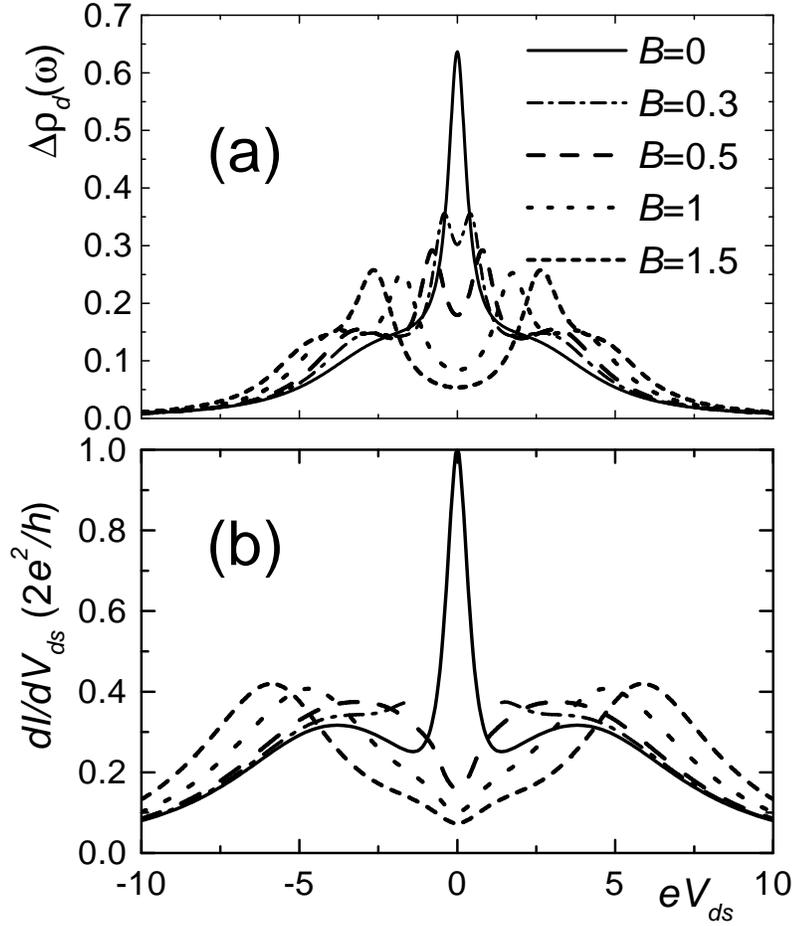,width=120mm,silent=} 
\caption{(a) Spectral density at the quantum dot as a function of energy
for $V_{ds}=0$ and (b) Differential conductance as a function of applied
voltage, for $T=0$, and different values of $B$. We remind the reader
that $T_{K}=0.45$.}
\label{s1b1}
\end{figure}

\begin{figure}
\hskip1.0cm\psfig{file=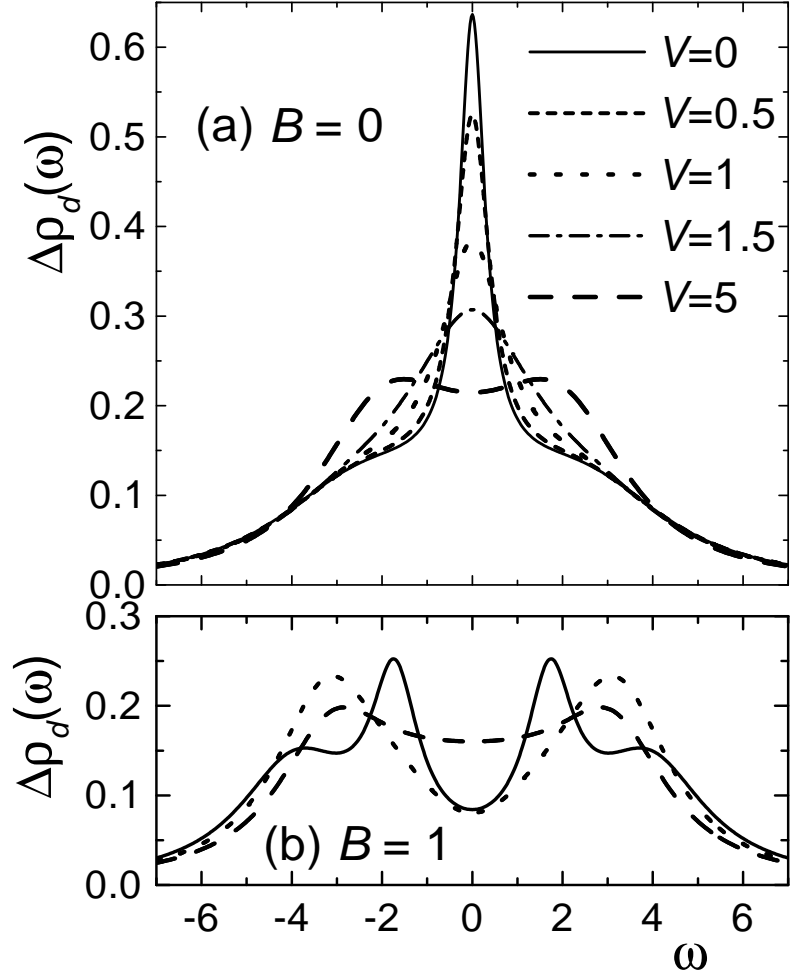,width=120mm,silent=} 
\caption{Spectral density as a function of energy for $T=0$, different
values of the applied voltage $V=eV_{ds}$ and (a) $B=0$, (b) $B=1$. 
Note that $T_{K}=0.45$.}
\end{figure}

\begin{figure}
\hskip1.0cm\psfig{file=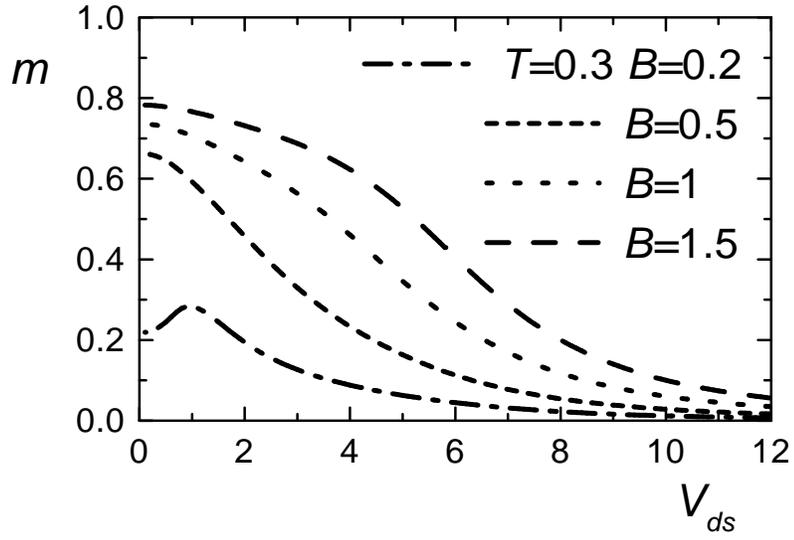,width=120mm,silent=} 
\caption{Magnetization as a function of the applied voltage for $T=0$
(except dashed dot line for which $T=2T_{K}/3=0.3$) and several values of $B.$}
\end{figure}

\begin{figure}
\hskip1.0cm\psfig{file=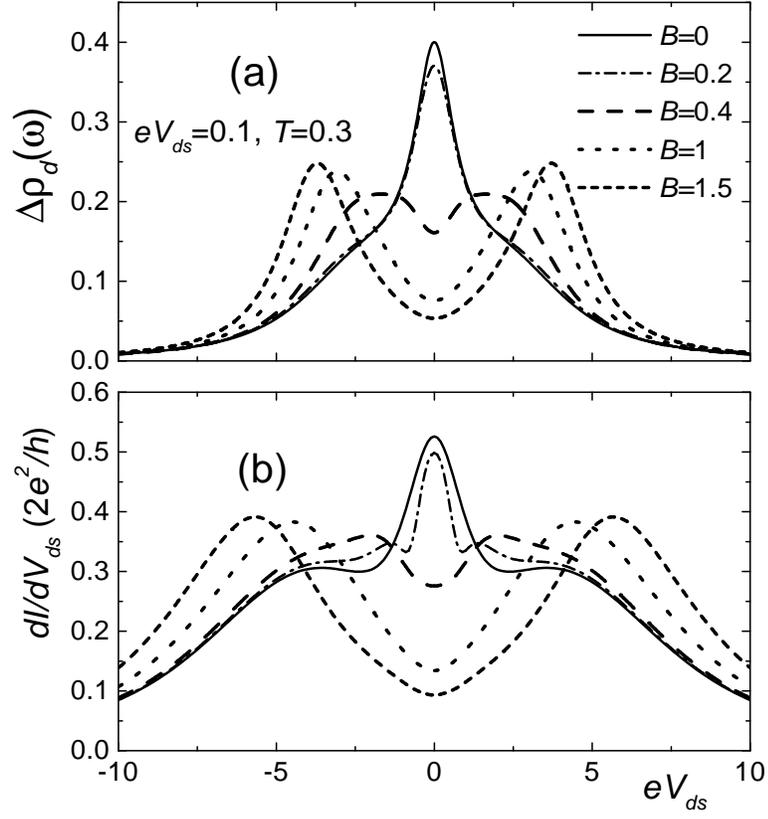,width=120mm,silent=} 
\caption{Spectral density a function of energy for $%
V_{ds}=0.1 $ and (b) Differential conductance as a function of applied
voltage for $T=2T_{K}/3=0.3$, and different values of $B$.}
\end{figure}

\begin{figure}
\hskip1.0cm\psfig{file=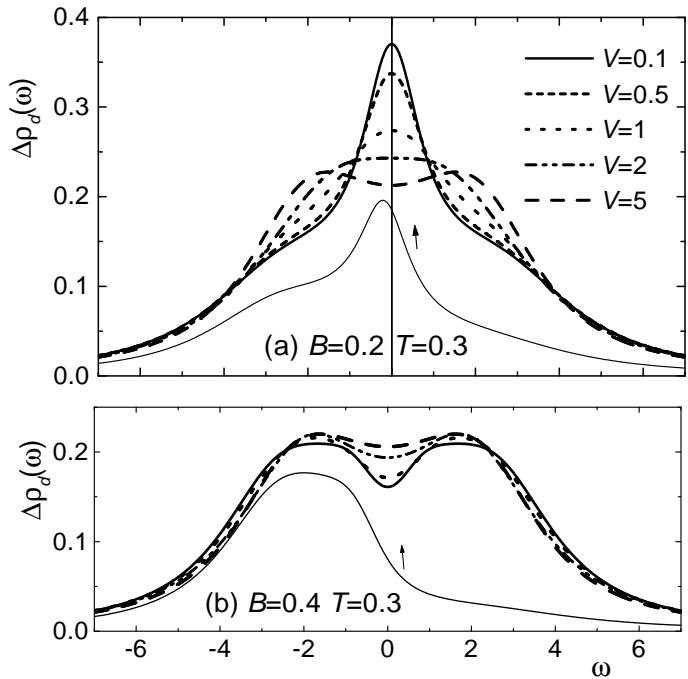,width=120mm,silent=} 
\caption{Spectral density as a function of energy for $T=2T_{K}/3=0.3$, different
values of the applied voltage $V=eV_{ds}$ and (a) $B=0.2$, (b) $B=0.4$. The
thin line corresponds to the contribution with spin up $\rho _{d\uparrow
}(\omega )$ for $V=0.1$.}
\end{figure}

\begin{figure}
\hskip1.0cm\psfig{file=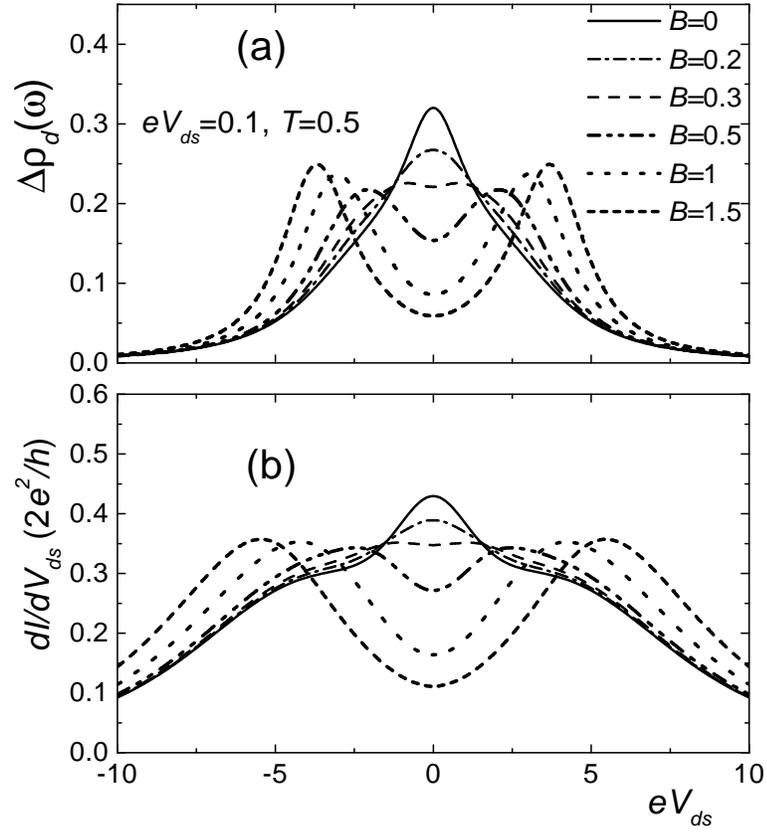,width=120mm,silent=} 
\caption{Same as Fig. 4 for $T=0.5 \sim T_{K}=0.45$.}
\end{figure}

\end{document}